\documentstyle[11pt,aaspp4]{article}

\newcommand{\be}{\begin{equation}}
\newcommand{\ba}{\begin{eqnarray}}
\newcommand{\ee}{\end{equation}}
\newcommand{\ea}{\end{eqnarray}}

\newcommand{\etal}{{\it et al. }}

\lefthead{Nakamura  }
\righthead{A Possible Interpretation of X-Ray afterglow of GRB980425}

\begin{document}
     

\title{Off-axis Emission from the Beamed Afterglow of Gamma Ray Bursts  and 
a Possible Interpretation of Slowly Declining X-Ray Afterglow of GRB980425}

\author{Takashi Nakamura}
\affil{Yukawa Institute for Theoretical Physics, Kyoto University,
Kyoto 606}

\received{}
\accepted{}

\begin{abstract}
A  formula for off-axis emissions from the beamed afterglow of GRBs in 
a synchrotron model is derived. The formula is applied to 
an inhomogeneous circumstellar matter obeying $n\propto r^{-2}$
suggested for SN1998bw/GRB980425. The  slow decline rate as well as
 the X-ray flux similar to the observed values  are obtained  for
appropriate choice of the parameters so that the association of GRB980425
with SN1998bw is  further strengthened.

(ApJ Letter in press)
\end{abstract}

\keywords{gamma rays: bursts---stars: supernovae}

\section{Introduction}
 Recently    the X-ray light curve of GRB980425 
and follow up data obtained by the BeppoSAX have been reported
(\cite{pian99}). The  NFI observation
identified an X-Ray source (S1) position-ally coincident
with the unusual Type Ic supernova SN1998bw (\cite{galama98,kul98}).
 Note here that the position of S1 is revised and  the other source S2 is
$\sim$ 4 $^{'}$ away from SN1998bw. 
The X-ray flux of S1 was $3\times 10^{-13}$erg s$^{-1}$cm$^{-2}$ in
April 26 and declined a factor of two  in six months. If this is 
an X-ray afterglow of GRB980425,  the decline rate ($\sim t^{-0.2}$) is 
extremely slower than usual, which needs an interpretation.
 If  GRB980425 is associated with SN1998bw,   the  distance is
$\sim 40$Mpc (\cite{galama98})  and the isotropic gamma ray
total energy   is unusually low $\sim 10^{48}$erg   compared
with  $>10^{51}$erg in  other GRBs for  which a  redshift measurement 
is available. 

 Such a low gamma ray luminosity might be interpreted if the beam direction  of
GRB980425 is different from the line of
sight(\cite{naka98,eich98}). In such an geometry we see only
scattered gamma rays so that the luminosity is low and the
energy of gamma rays is not high $< 500$keV (\cite{naka98,eich98}). 
This also interprets the fact
that GRB980425  belongs to so called Non High  Energy class of GRBs (\cite{pen97}). 
It is  shown that if the angle between the beam and the line of sight  is
$\sim 30^{\circ}$, the luminosity can be  $\sim 10^{-3}$ smaller 
than usual (\cite{eich98}). Such a beam model is relevant now since
an evidence for the  beaming  has recently been suggested
from the rapid decline of optical afterglow of GRB990123 after day 2
(\cite{kul99}) and the rapid decline of GRB980519(\cite{halp99}).

An angle of $\sim 30^{\circ}$ has also been suggested  to
interpret the linear optical polarization ($\sim$0.5\% in June $20\sim 
23;  $ \cite{kay98}) of SN1998bw   .
If the supernova ejecta is an oblate or a prolate shape with an aspect
ratio 1 : 2 $\sim$ 3 and the line of sight is $\sim 30^{\circ}$ away from the 
symmetry axis, the observed linear polarization degree  might be
explained (\cite{hoef98}).

From the radio  observation of SN1998bw  it is suggested that 
the density distribution of the circumstellar matter follows
$n=n_0r_0^2/r^2$ with $n_0r_0^2\sim 10^{34}/$cm,
 which is consistent with the constant mass loss rate in the
progenitor(\cite{li99}). It is  also claimed that the energy of the
order of $10^{50}$erg should be injected twice with the injection
velocity of the order of 0.5c, which strengthens the link between
SN1998bw  and GRB980425.

In this Letter we try to interpret the slowly declining X-ray afterglow of
GRB980425 in the beam model of GRBs assuming the line of sight
is  $\sim 30^{\circ}$  away from the axis of the beam.  We  adopt
the density distribution of the circumstellar matter  suggested by
\cite{li99}. We will obtain
the  decline rate and the X-ray flux similar to the observed values
for appropriate choice of the parameters so that the association of GRB980425
with SN1998bw will  be further strengthened.

\section{Off-axis emission  from  the beamed afterglow  } 
\cite{granot98} as well as \cite{wood99}  derived a general formula 
 to compute the
off-axis emission from beamed GRBs. Here we adopt their formulations
 and notations. Let us use a spherical coordinate system ${\bf r}=(r,
\theta,\phi)$  where the coordinate are measured in the lab frame; let
 the $\theta =0$ axis ($z$-axis) points to the detector and ${\bf
r}=0$ be the central engine. Let  also   $D$ be the distance
to the source and $\alpha=r\sin \theta/D$ be the angle that a
given ray makes with the normal to the detector.  Then the observed 
 flux is given by
\be
F_\nu(T) = \frac{\nu D}{\gamma\beta} \int_0^{2\pi} d\phi 
\int_0^{\alpha_m}
\alpha^2 d\alpha \int_{\nu\gamma(1-\beta)}^{\nu\gamma(1+\beta)}
\frac{d\nu'}{{\nu'}^2}
~\frac{j'_{\nu'}[\Omega' ,{\bf r},T+\frac{r\mu}{ c}]}
{\{1-\mu^2\}^{3/2}},
\ee
where $\mu = (1-\nu'/\gamma\nu)/\beta$. $\alpha_m$, $T$ and $j'_{\nu'}$ are the maximum value of $\alpha$,
 the arrival time of a photon at the detector and the rest frame
emissivity measured in ${\rm erg~s^{-1}~cm^{-3}~Hz^{-1}~sr^{-1}}$,
respectively. Note here that $'$ means the physical quantity in the
rest frame.

We adopt the standard fire ball model(\cite{piran98}).
As an emissivity we extend a simple synchrotron model( \cite{sari98})
used in a homogeneous ambient gas  to the inhomogeneous  circumstellar
matter following $n=n_0(r/r_0)^{-d} $. 
d=2 corresponds to the constant mass loss
from the progenitor of GRBs. This kind of a model has ever been studied in
a different context (\cite{mes98}).

  We assume that electrons are accelerated
in the shock to a power-law distribution of Lorenz factor $\gamma_e$
with a minimum  Lorenz factor   $\gamma_m$: $N(\gamma_e)\propto
\gamma_e^{-p}d\gamma_e, \gamma_e > \gamma_m$. $\gamma_m=G\gamma$ where
$\gamma$ is the Lorenz factor of the shocked fluid and
$G=\epsilon_e(p-2)/(p-1)m_p/m_e$  with $\epsilon_e$ being the efficiency of the
acceleration.  Since $p=2\sim 2.5$  is suggested from observations,  $G=
60(\epsilon_e/0.1)(3(p-2)/(p-1)) $.
The magnetic field strength is given as
$   B =B_0\gamma ({r}/{r_0})^{-d/2}$ where
  $ B_0=\sqrt{32\pi m_pn_0\epsilon_B}$.
 $\epsilon_B$ measures the ratio of  the magnetic
field energy to the total thermal energy. 
The rest frame radiation power ($P'$) and the characteristic synchrotron
frequency ($\nu'(\gamma_e)$)  from a randomly oriented electron with
Lorenz factor $\gamma_e \gg 1$ in a magnetic field $B$ are given by
$P'= P_0 \gamma ^2\gamma_e^2({r}/{r_0})^{-d}$
and $\nu'(\gamma_e)=\nu_0 \gamma \gamma_e^2({r}/{r_0})^{-d/2}$
where $P_0={\sigma_TB_0^2c}/{6\pi}$
and
$\nu_0={eB_0}/{2\pi m_ec}=3.5\times 10^5{\rm Hz}({n_0}/{1{\rm cm^{-3}}})^{1/2}
({\epsilon_B}/{0.1})^{1/2}$.

The peak spectral power flux does not depend on $\gamma_e$
and  occurs at $\nu'(\gamma_e)$ and is given by
$P'_{\nu',{\rm max}}={P'}/{\nu'}={P_0\gamma}/{\nu_0}({r}/{r_0})^{-d/2}$. 
  The critical gamma factor $\gamma_c$ is defined by
  $ \gamma_cm_ec^2=P'(\gamma_c)\tau=P'(\gamma_c)r/(\gamma c)$
and is given by
$\gamma_c={F}/{\gamma}({r}/{r_0})^{d-1}$ where
 $F={m_ec^3}/{P_0r_0}$.
 Then the energy spectra depend on two frequencies  defined by 
$\nu'_m\equiv\nu'(\gamma_m)=\nu_0G^2\gamma^3({r}/{r_0})^{-d/2}$
and
 $\nu'_c\equiv\nu'(\gamma_c)=\nu_0F^2\gamma^{-1}({r}/{r_0})^{3d/2-2}$
(\cite{sari98}) .

There are two different cases in the energy spectra depending on 
whether $\nu'_m > \nu'_c$ ({\it fast cooling}) or
$\nu'_m < \nu'_c$ ({\it slow cooling});(\cite{sari98}).  In both cases
the emissivity  $j'_{\nu'}$ is expressed in the form as
\be
 j'_{\nu'}=P'_{\nu',{\rm max}}
\frac{\gamma n_0}{4\pi}(\frac{r}{r_0})^{-d}f(\nu'),
\ee
where $f(\nu')$ is given for {\it fast cooling} case as
\[ f(\nu') = \left\{
\begin{array}{@{\,}ll}
(\nu'/\nu'_c)^{1/3} & \nu'_c > \nu' \\
(\nu'/\nu'_c)^{-1/2} & \nu'_m > \nu' > \nu'_c \\
(\nu'_m/\nu'_c)^{-1/2}(\nu'/\nu'_m)^{-p/2}  & \nu' > \nu'_m, 
\end{array}
\right. \]
while for {\it slow cooling} case we have a smilar formula(\cite{sari98}).

Let $\Delta \theta$ and $\theta_v$ be the beaming half-angle and
the angle between the direction to the detector and the axis
of the emission cone. In case of  $\theta_v=0$, Eq. (1) is evaluated as
\be
F_\nu(T) = \frac{r^3n_0}{3D^2\beta} (\frac{r}{r_0})^{-d} I 
; ~~ I=\int_{\nu\gamma(1-\beta)}^{\nu\gamma(1-\beta\cos\Delta\theta)}
\frac{\nu d\nu'}{2{\nu'}^2}P'_{\nu',{\rm max}}f(\nu').
\ee
For $\gamma  \gg 1$ and $\Delta \theta < 1$ , the integral is expressed as
\be
I=\int_{1}^{\gamma^2\Delta\theta^2}
\frac{ ds}{s^2}\gamma P'_{\nu',{\rm max}}f(\nu\gamma s/2).
\ee 
If $\gamma\Delta\theta \gg 1$ ,
$I$ does not depend on $\Delta \theta$. This is expected since in this
case the detector detects the radiation only from the half-angle of
$\gamma ^{-1}$ and can not observe the edge of the beam. For $d=0$, Eq. 
(4) gives  the same results as in \cite{sari98}.
\footnote{In reality there is 10\% or so difference depending on the
parameter $p$. However such a difference  is consistent 
in  a simple model of emissivity in \cite{sari98}.}

Now let us consider another extreme case of $1 > \theta_v > \Delta
\theta > \gamma ^{-1}$. In this case the line of sight from ${\bf
r}=0$ does not pass through the beam.  Then Eq. (1) is evaluated as
\be
F_\nu(T) = \frac{r_0^3n_0P_0}{12\pi D^2\nu_0}
 (\frac{r}{r_0})^{3(2-d)/2} I ;
 ~~I=\int_{\nu\gamma(1-\beta\cos(\theta_v-\Delta \theta) )}
^{\nu\gamma(1-\beta\cos(\theta_v+\Delta \theta))}
2\phi_v\frac{\gamma \nu d\nu'}{{\nu'}^2}f(\nu'),
\ee
where
$\cos \phi_v =(\cos
\Delta\theta-\cos\theta\cos\theta_v)/{\sin\theta_v\sin \theta}.$

Eq.(5) is a general formula for off-axis emission from beamed GRBs in 
a synchrotron model. In the next section we shall apply the
formula to GRB980425.

\section{Application to GRB980425}
Now let us consider d=2 case which is suggested from SN1998bw (\cite{li99}).
Then Eq. (5) is rewritten as
\be
F_\nu(T) = \frac{r_0^3n_0P_0}{12\pi D^2\nu_0}I; ~~I=\int_{(1-\beta\cos(\theta_v-\Delta \theta) )}
^{(1-\beta\cos(\theta_v+\Delta \theta))}
2\phi_v\frac{ds}{s^2}f(\nu\gamma s).
\ee

The position of the shock is related to the arrival time $T$ as
$r=cT/(1-\beta \cos \theta_v)\sim 1.9\times 10^{16}{\rm cm}
({\theta_v}/{30^{\circ}})^{-2}T_d$
 where  $T_d$ is the arrival time in days.
 $\gamma$ is given for the adiabatic case  by
\be
\gamma = (\frac{E_b}{m_pc^2n_0r_0^2r2\pi\Delta\theta^2})^{1/2}
= 63 (\frac{E_b}{{\rm 10^{51}erg}})^{1/2}(\frac{1-\beta \cos
\theta_v}{\Delta\theta^2})^{1/2} 
(\frac{n_0}{1{\rm cm^{-3}}})^{-1/2}
(\frac{r_0}{10^{17}{\rm cm}})^{-1}T_d^{-1/2},
\ee
where $E_b$ is the total energy in the beam.
 
The two characteristic energy $h\nu_m$ and $h\nu_c$ at the detector
  are given by
\ba
&&h\nu_m=2\times 10^{-2}{\rm
eV}(\frac{G}{60})^{2}(\frac{1-\beta \cos
\theta_v}{0.1})^{-1}(\frac{\gamma}{20})^2
(\frac{r}{r_0})^{-1}(\frac{\epsilon_B}{0.1})^{1/2}(\frac{n_0}{1{\rm cm^{-3}}})^{1/2}\\
&&h\nu_c=8.7\times 10^{-3}{\rm
eV}(\frac{1-\beta \cos
\theta_v}{0.1})^{-1}(\frac{\gamma}{20})^{-2}(\frac{r}{r_0})
(\frac{\epsilon_B}{0.1})^{-3/2}(\frac{n_0}{1{\rm cm^{-3}}})^{-3/2}
(\frac{r_0}{ 10^{17}{\rm cm}})^{-2}
\ea

Now consider the observed radiation in optical and X-ray bands.
$I$ is evaluated for $\theta_v \sim30^{\circ}$ and {\it fast cooling} case as
\be
I=2^{4+p/2}\frac{1}{\theta_v^{2+p}}(\frac{\Delta\theta}{\theta_v})^2
G^{p-1}(\frac{r}{r_0})^{1-p/2}\gamma^{p-2}F(\frac{\nu_0}{\nu})^{p/2}.
\ee
Then the energy flux in a band of   $\nu_d < \nu  < \nu_u$ is given by
\ba
&&F_b\equiv \int_{\nu_d }^{\nu_u}F_{\nu}d\nu =1.74 \times 10^{-12}
{\rm erg~s^{-1}~cm^{-2}}
 C_p\frac{2}{p-2}
((\frac{\nu_0}{\nu_d})^{p/2-1}-(\frac{\nu_0}{\nu_u})^{p/2-1})\nonumber \\
&&(\frac{r_0}{ 10^{17}{\rm cm}})^{4-p}(\frac{n_0}{1{\rm cm^{-3}}})^{2-p/2}
(\frac{D}{40{\rm Mpc}})^{-2}(\frac{\theta_v}{30^{\circ}})^{-4}(2\frac{\Delta\theta}{\theta_v})^{4-p}(\frac{E_b}{{\rm
10^{51}erg}})^{p/2-1}T_d^{2-p},
\ea
where $C_p=2^{3(p-2)/2}12140^{(p-2)}({G}/{60})^{p-1}$
Let us compute $F_b$ in X-ray band ($\nu_d=1.6$keV, $\nu_u=10$keV)
 and V band ($\nu_d=5\times 10^{14}$Hz, $\nu_u=6\times 10^{14}$Hz)
 for $p=2.2$  and $p=2.5$.

For $p=2.2$
\ba
&&F_b=
6.5 \times 10^{-13}{\rm erg~s^{-1}~cm^{-2}}(
1.4 \times 10^{-13}{\rm erg~s^{-1}~cm^{-2};~~V~ band})(\frac{r_0}
{ 10^{17}{\rm cm}})^{1.8}\nonumber \\
&& (\frac{n_0}{1{\rm cm^{-3}}})^{0.95}(\frac{D}{40{\rm Mpc}})^{-2}
(\frac{\theta_v}{30^{\circ}})^{-4}(\frac{\epsilon_e}{0.1})^{1.2}
(\frac{\epsilon_B}{0.1})^{0.05}(2\frac{\Delta\theta}{\theta_v})^{1.8}
(\frac{E_b}{{\rm10^{51}erg}})^{0.1}T_d^{-0.2}.
\ea
In this case the declining rate   is almost the same as the observed
value while the absolute value of the X-ray flux is similar to 
the observed one and can be adjusted by choosing the appropriate
values of various parameters such as
$\theta_v, \Delta\theta,\epsilon_e, \epsilon_B,r_0,n_0$ and $E_b$.

For p=2.5
\ba
&&F_b=
8.0 \times 10^{-13}{\rm erg~s^{-1}~cm^{-2} }
(3.0 \times 10^{-13}{\rm erg~s^{-1}~cm^{-2};~~V~ band })
 (\frac{r_0}{ 10^{17}{\rm cm}})^{1.5}\nonumber\\
&&(\frac{n_0}{1{\rm cm^{-3}}})^{0.875}(\frac{D}{40{\rm Mpc}})^{-2}
(\frac{\theta_v}{30^{\circ}})^{-4}(\frac{\epsilon_e}{0.1})^{1.5}
(\frac{\epsilon_B}{0.1})^{0.125}(2\frac{\Delta\theta}{\theta_v})^{1.5}
(\frac{E_b}{{\rm10^{51}erg}})^{0.25}T_d^{-0.5}.
\ea
In this case, the amplitude of the X-ray flux can be similar to the observed
value but the decline rate is somewhat too rapid.

For comparison, we shall compute the energy flux for an aligned case
($\theta_v=0$). Since  in this case the position of the shock is related to the
arrival time of the radiation  by $r=2\gamma^2cT$, $\gamma$ is given
by
\be
\gamma = 13 (\frac{E_b}{{\rm 10^{51}erg}})^{1/4}(\frac{\Delta\theta}{15^\circ})^{-1/2} (\frac{n_0}{1{\rm cm^{-3}}})^{-1/4}
(\frac{r_0}{10^{17}{\rm cm}})^{-1/2}T_d^{-1/4}
\ee

$F_b$ is given for  $p=2.2$ as
\ba
&&F_b=
7.8 \times 10^{-11}{\rm erg~s^{-1}~cm^{-2} }
(1.7 \times 10^{-11}{\rm erg~s^{-1}~cm^{-2};~~V~ band })
 (\frac{r_0}{ 10^{17}{\rm cm}})^{0.5}\nonumber\\
&&(\frac{n_0}{1{\rm cm^{-3}}})^{0.3}(\frac{D}{40{\rm Mpc}})^{-2}
(\frac{\Delta\theta}{15^{\circ}})^{-1.6}(\frac{\epsilon_e}{0.1})^{1.2}
(\frac{\epsilon_B}{0.1})^{0.1}(\frac{E_b}{{\rm10^{51}erg}})^{0.8}T_d^{-0.9}.
\ea
while for $p=2.5$
\ba 
&&F_b=
2.7 \times 10^{-11}{\rm erg~s^{-1}~cm^{-2} }
(1.7 \times 10^{-11}{\rm erg~s^{-1}~cm^{-2};~~V~ band })
 (\frac{r_0}{ 10^{17}{\rm cm}})^{0.5}\nonumber\\
&&(\frac{n_0}{1{\rm cm^{-3}}})^{0.375}(\frac{D}{40{\rm Mpc}})^{-2}
(\frac{\Delta\theta}{15^{\circ}})^{-1.75}(\frac{\epsilon_e}{0.1})^{1.5}
(\frac{\epsilon_B}{0.1})^{0.125}(\frac{E_b}{{\rm10^{51}erg}})^{0.875}
T_d^{-1.125}.
\ea

Let us apply  Eq. (16) to GRB990123. There was a break in the optical
light curve at   $T_d=2$. This may be due to the widening of the beaming half-angle 
(\cite{rhoads97,kul99}). From the condition of $\gamma =\Delta \theta ^{-1}$ at $T_d=2$, $\Delta\theta$ is expressed by the other parameters as
\be
\Delta\theta=1.8^{\circ}(\frac{E_b}{{\rm10^{51}erg}})^{-1/2}(\frac{r_0}
{ 10^{17}{\rm cm}}) (\frac{n_0}{1{\rm cm^{-3}}})^{0.5}
\ee
Adopting $D=$10Gpc as the luminosity distance to GRB990123 
($z=1.6$ ; \cite{bloom99})
 we have
\ba
&&F_b=
1.8 \times 10^{-14}{\rm erg~s^{-1}~cm^{-2} }
(1.1 \times 10^{-14}{\rm erg~s^{-1}~cm^{-2};~~V~ band })
 (\frac{r_0}{ 10^{17}{\rm cm}})^{-1.125}\nonumber \\
&&(\frac{n_0}{1{\rm cm^{-3}}})^{-0.5}(\frac{D}{10{\rm Gpc}})^{-2}
(\frac{\epsilon_e}{0.1})^{1.5}(\frac{\epsilon_B}{0.1})^{0.125}
(\frac{E_b}{{\rm10^{51}erg}})^{1.75}T_d^{-1.125}.
\ea
For example, if $E_b\sim{{\rm10^{52}erg}}$ with the other parameters
unchanged, the decline rate and amplitude of the flux in X-ray  and
the optical bands are similar to the observed values. However even if
$E_b\sim{\rm 10^{51}erg}$, by appropriate choice of the other parameters
such as $r_0, n_0,\epsilon_e$ and $\epsilon_B$, we may also have the observed 
amplitude.

\section{Discussion}
   Since a possible association of GRBs with Type Ib/Ic supernovae is
suggested only for GRB980425, we had better wait for more
events to confirm the conjecture.
However recently another possible association of GRB with the supernova is
suggested for GRB980326.  The decline of the afterglow was leveled off 
at $\sim$20 days after the GRB, which was considered as the  host
galaxy. 
Keck observed the host galaxy nine months after the burst again  but
could not find the host galaxy. A possible explanation for this
peculiar event  is that the apparent
level off at $\sim$20 days is due to the peak of supernova like  SN1998bw(\cite{bloom99b}).
If this is the case, at least two examples of association of GRBs with  supernovae
exist.

For GRB980425 in our model the break  in the light curve
when $\gamma=\theta^{-1}$ should occur  at
\be
T_d=544 (\frac{E_b}{{\rm 10^{51}erg}})(\frac{\theta_v}{30^{\circ}})^{2}
(\frac{n_0}{1{\rm cm^{-3}}})^{-1}
(\frac{r_0}{10^{17}{\rm cm}})^{-2}.
\ee 
The break time does not depend on $\Delta\theta$ but on
$E_b,\theta_v,  n_0$ and $r_0$. Before the break time, Eq.(12) and
(13) can be used so that  the optical luminosity is much lower than the
luminosity of SN1998bw for $T_d < 180$ (\cite{iwa98}). However the decline rate is so low that the
luminosity from the afterglow might overcome the  luminosity of the
supernova and the host galaxy  before $T_d\sim 544$. Therefore it is 
important to observe  SN1998bw up to $\sim 2$ years after the supernova event. 
 
In conclusion, if the line of sight
is  $\sim 30^{\circ}$  away from the axis of the beam for GRB980425,
the  decline rate and the X-ray flux similar to the observed values
are obtained  for appropriate choice of the parameters. This strengthens  the link between GRB980425
and  SN1998bw.

\acknowledgments
 This work was supported in part by
Grant-in-Aid of Scientific Research of the Ministry of Education,
Culture, and Sports, No.11640274 and 09NP0801.

\end{document}